\documentclass[a4paper,12pt]{article}

\usepackage{amsmath,amssymb}

\usepackage{amsmath}
\usepackage[mathscr]{eucal}

\usepackage{lmodern}
\usepackage{extarrows}
\usepackage{rotating}
\usepackage{MnSymbol}
\usepackage{caption}
\usepackage{float}

\begin{document}

\begin{center}

\section*{Quantum vortices in entanglement: \\ a novel idea for large vortex filaments.  }

\vspace{3mm}

{S.V. Talalov}

\vspace{3mm}

{\small Institute of Digital Technology, Togliatti State University, \\ 14 Belorusskaya str.,
 Tolyatti, Samara region, 445020 Russia.\\
svt\_19@mail.ru}

\begin{abstract}
In this study, we propose a new approach to describing certain macroscopic   objects
 that can arise in a quantum fluid.  These objects are  formed by means of quantum entanglement from  the circular-shaped mesoscale and microscale  vortices, and  can be interpreted as a vortex filaments with any  shape and size.  The method is based on a quantization scheme for classical closed vortex filaments that was proposed by the author early \cite{Tal18,Tal22_1,Tal_Chaos25J}.  The model we consider examines the instantaneous picture of the locations in space $\mathbb R_3$  of such filaments with a small, but non-zero, core diameter. 
 Both  energy  and  circulation of the studied  filaments are calculated using the proposed approach.
We demonstrate that the adopted concept leads to the emergence of secondary vortices around these investigated filament-like objects.
 We also study the specific mechanisms by which large vortex loops can disconnect and create the filament fragments . 
From our point of view, the proposed approach to describing vortex filaments within a vortex tangle can be seen as an important step toward understanding the appearance of quantum turbulence.
\end {abstract}

\vspace{5mm}

{\bf keywords:} ~   quantum vortices,  ~quantum turbulence.

\end{center}

%\vspace{5mm}

 {\bf PACS numbers:}   47.10.Df    ~~47.32.C

\vspace{5mm}

\section{Introduction}

\paragraph~

~~~The complexity of describing turbulence is best summed up by Heisenberg's famous saying about two questions for God: why relativity, and why turbulence. ''I realy think He may have an answer to the first question'' \footnote{The quote of this Heisenberg's statement is taken from the book \cite{Glaik}.}
Of course, Heisenberg paid a lot of attention to this problem himself (see, for example, \cite{Heis}). However, it would be an overstatement to say that this phenomenon has been fully understood at this point. This poor understanding is especially relevant when we want to provide a quantitative description of phenomena and processes that occur here.
It seems that the current  quality  description  of turbulence  is that it is a complex tangle of vortices \cite{Feyn_2}.
Since then, a large number of articles have been written on this approach. However, the author does not attempt  to provide a comprehensive review of all of them here. Most of this research seems to have been summarized in books (see, for example, \cite{Vinen,Tsu_1,Sonin,Nemir}).

Thus, the problem of describing quantum turbulence necessarily involves the problem of vortex quantization.   The most common point of view is that quantum vortices are topological defects in quantum fluids. This hypothesis does not contradict some specific  experiments with superfluid helium.  However, here we would like to draw attention to Feynman's remark \cite{Feyn_1}  that quantum fluid dynamics and the theory of super-fluidity are not the same. This, combined with difficulties in providing quantitative descriptions of turbulent phenomena, stimulates the development of new approaches to solving these problems.

In a number of recent papers, the author has suggested a new approach to quantization of the vortex loops (see, for example, \cite{Tal18,Tal22_1,Tal_PoF23,Tal_Chaos25J}).
The central elements of the suggested approach are:
\begin{itemize}
\item A quantum vortex is a quantized classical dynamical system, but it is not a topological defect in a quantum fluid. We quantize  a filament with a small, but non-zero core radius ${\sf a}$. This filament evolves according to the Local Induction Approximation;
\item The dynamics of a thin vortex filament is described both in terms of geometrical variables and non-geometrical ones.  To include the movement of surrounding fluid, we declare circulation $\Gamma$ as an additional independent dynamic variable; 
\item We redefine the set of dynamic variables, so that it is possible to take the following steps:
\begin{itemize}
\item Extension of the space-time symmetry group of the theory to the centrally extended Galilei group $\widetilde{\mathcal G}_3$, already at the classical level;
\item Using the formalism of many-body physics  to describe vortex interactions;
\end{itemize}
\item We apply a group-theoretic approach to define  the full energy of a thin  (${\sf a} \to 0$, ${\sf a} \not= 0$ ) vortex filament. Indeed, Lee algebra of the  group $\widetilde{\mathcal G}_3$ has three Cazimir functions:  
		 \[ {\hat C}_1 = \mu_0 {\hat I}\,,\quad 
  {\hat C}_2 = \left({\hat M}_i  - \sum_{k,j=x,y,z}\epsilon_{ijk}{\hat P}_j {\hat B}_k\right)^{\!2} 
  \quad {\hat C}_3 = \hat H -  \frac{1}{2\mu_0}\sum_{i=x,y,z}{\hat P}_i^{\,2}\,,\]                        
       where        ${\hat I}$ is the unit operator,     ${\hat M}_i$,   $\hat H$,  ${\hat P}_i$         and  ${\hat B}_i$  ($i = x,y,z$)
        are the respective generators of rotations, time and space translations and Galilean boosts, value $\mu_0$ is a central charge. 
		As usual, the function  ${\hat C}_3 $  is interpreted as  an  ''internal energy of the particle'' as well as the central charge $\mu_0$ is  interpreted as a ''mass''.  This means that we can define vortex energy as follows\footnote{It is appropriate to recall Donnelly's remark here: {\it''\dots  considering how small the vortex core in helium II is, i.e., of order an angstrom, it would seem that one either ought to know how it is constructed or one ought to find a way to ignore it. Unfortunately neither goal has been achieved''} \cite{Donn}.	}
		\begin{equation}
		\label{E_general}
 {\cal E}_{cl} ~=~ \frac{{\bf p}^{\,2}}{2\mu_0} ~+~ {\hat C}_3(\varpi, \chi, \dots)\,,
\end{equation}
where ${\bf p}$ is vortex momentum and the variables $\varpi, \chi, \dots$ some set of the ''internal'' variables;
\item We consider only round-shaped micro- and meso-sized loops with some flow in the core.  Additionally, we consider the small oscillations of these loops, known as Kelvin waves;
\item  The spectra of  the circulation $\Gamma$ as well as the energy ${\cal E}$  are eigenvalues of certain spectral problems in Hilbert space;
\item We construct Fock space to describe a  system of the interacting vortices.
\end{itemize}

The model that describes the turbulence at an initial stage was developed in the paper 
\cite{Tal_PoF23}. The approach suggested considers the creation and subsequent separation of circular micro-and meso-vortices. In general, the process of transformation of $m$ microvortices into $n$ micro-vortices has been described.
But the vortex tangle in turbulent flow consists of filaments of arbitrary shapes and sizes. Unlike micro-vortices, these non-local objects cannot appear in the form of some kind of local fluctuations in a laminar flow.
Therefore, the question of how large vortex filaments arise in a vortex tangle remains interesting. In this paper, we consider the possible mechanism of the appearance of these objects.

\section{The base structures of the classical theory}

~~~The main points of classical description of vortex loops, including the choice of variables and Hamiltonian structure, are presented in detail by the author in the cited papers.Here we present the key points concisely, avoiding excessive repetition of the results that have been published earlier.

 As the first step, we fix the fundamental and auxiliary constants of the theory. The fundamental  constants  are as follows: 1) fluid density $\varrho_0$, 2) the speed of sound in a fluid $v_0$, 3) length constant $R_f$, 4) mass constant $\mu_0$.
  The constant  $R_f$ may be related to the size of a fluid molecule, inter-atomic distance or  size of a stable molecular cluster.  The mass parameter $\mu_0$ is a central charge for the central extended Galilei group.     Along with the fundamental constants, we will also use auxiliary constants.
Firstly, we will use some length scale $L$. This constant can be fixed one way or the other, depending on the specific physical problem.
	For example, we can define this constant as $L \simeq \max|{\boldsymbol r}_1 - {\boldsymbol r}_2|$, where  ${\boldsymbol r}_1,  {\boldsymbol r}_2 \in D$ for the case where vortices evolve within a bounded domain $D$. When we consider the vortex filament in a tangle, we can assume $L = {\cal L}^{-1/2}$, where symbol  ${\cal L}$ means the vortex filament density \cite{Vinen,Nemir}. 
	  At this stage of our study, we will not provide an exact definition of the constant $L$. 	
	Secondly, we introduce the notations
	\[  t_0 = \frac{L}{v_0}\,, \qquad    {\mathcal E}_0 = \mu_0 v_0^2\,, \qquad
	\tilde\mu_0 = \pi\varrho_0 R_f^3\]
to simplify the formulas. 
The  constants $t_0$ and  ${\mathcal E}_0$ define the  time and energy scales in constructed classical  theory.

We use the Local Induction Approximation to describe thin vortex filaments in a certain flow in their core. As geometrical object, this filament is described by the radius vector $\boldsymbol r(t, {\sf s})$, where $t$ is physical time and ${\sf s}$ is the natural parameter along the curve.

At the initial stage, we only consider loops that have a circular shape with a radius of $R$.  Exploring the dynamics of the vortex ring, it will be more convenient  to use dimensionless parameters $\tau$ and $\xi$ instead of the ''real time'' $t$ and the natural parameter ${\sf s}$:
\begin{equation}
        \label{tau_s}
 \tau    ~=~   \frac{t\Gamma  }{4\pi R^2}\,,	\qquad \quad \xi ~=~ \frac{{\sf s}}{R}\,, \qquad
\xi \in [0,2\pi]\,.
 \end{equation}
Thus, we  consider the following equation for a projective vector 
${\mathfrak r} =  {\boldsymbol{r}}/R$:

\begin{eqnarray}
        \label{LIE_pert}
        \partial_\tau {\mathfrak r}(\tau ,\xi)  & = &
        \alpha\, \Bigl(\partial_\xi{\mathfrak r}(\tau ,\xi)\times\partial_\xi^{\,2}{\mathfrak r}(\tau ,\xi)\Bigr)   ~+ \nonumber \\
				~~ & + & \omega\,\Bigl(2\,\partial_\xi^{\,3}{\mathfrak r}(\tau ,\xi) ~+~ 
        {3}\,\bigl\vert\, \partial_\xi^{\,2}{\mathfrak r}(\tau ,\xi)\bigr\vert^{\,2}\partial_\xi{\mathfrak r}(\tau ,\xi)\Bigr)\,,
				        \end{eqnarray}
where the values $\alpha$   and  $\omega$ are finite dimensionless constants.
In total,  parameter $\omega$  is determined by the velocity of fluid flow  $\Phi_\omega$ within the vortex core.  A detailed deducing  of this equation { as well as deducing the explicit expression of the parameter $\omega$  from the conventional physical values} was made in  the book \cite{AlKuOk}.

The  Eq.(\ref{LIE_pert}) has the exact solution

\begin{equation}
        \label{our_sol}
 {\mathfrak r}(\tau ,\xi) ~=~ \Bigl(\, \frac{q_x}{R} +  \cos(\xi +\phi)\,,\quad \frac{q_y}{R} +  \sin(\xi +\phi  )\,, \quad \frac{q_z}{R} + \alpha \tau \,\Bigr)\,, 
\end{equation}
where $\phi \equiv \phi(\tau) =  \phi_0 +  \omega\tau$. 
This variable defines the rotation of the circular curve (\ref {our_sol}), and consequently models the flow into the core of the vortex.
Vector $\boldsymbol{q} =
 (q_x, q_y, q_z)$ defines the position of the curve in the coordinate system. 
 Later in this paper, we will be looking into this particular solution.

 In our approach, we consider the radius of the vortex $R$ to be a dynamical variable.
For subsequent purposes, we will use the variable $\Delta R$ instead of the value $R$:
\begin{equation}
        \label{Delta_R}
	\Delta R ~=~ \sqrt{R^2 - R_f^2}\,,
				\end{equation}

Let the vector ${\boldsymbol b}$ means the binormal unit vector of the considered circular filament.
Thus, the variables ${\cal A} = \{ \Gamma, {\boldsymbol q},\Delta R, \phi(\tau), {\boldsymbol b}\}$ describe the considered dynamical system which  is associated  with solution (\ref{our_sol}).
Futher, we transform the set ${\cal A} \to {\cal A}^{\,\prime} $  for subsequent quantization purposes.
To define new variables, we use  the conventional formula for the canonical momentum ${\boldsymbol p}$  \cite{Batche}:
	  
   \begin{equation}
        \label{p_can}
        {\boldsymbol p} ~=~ {\boldsymbol p}_0 ~+~
				\frac{\varrho_0}{2 }\,\int\,\boldsymbol{r}\times\boldsymbol{w}(\boldsymbol{r})\,dV\,.
        \end{equation}             
   The summand ${\boldsymbol p}_0$ takes into account the  momentum which arises when the surrounding fluid has a non-zero velocity $\boldsymbol{v}$ without vortex motion\footnote{ In this case, this term can contains the summand 	$ {\sf M}\boldsymbol{v}$, where the value ${\sf M}$ means some ''mass''. The term ${\boldsymbol p}_0$ may be important, for example, when we  consider Galilean transformations. }. To simplify our final formulas, we assume that  ${\boldsymbol p}_0 = 0$ in our subsequent  studies. The second summand is the  vortex canonical momentum ${\boldsymbol p}$  \cite{Batche}.   		
		  The vorticity  ${\boldsymbol{w}}(\boldsymbol{r})$  of the  closed vortex filament is calculated as follows
   \begin{equation}
        \label{vort_w}
     {\boldsymbol{w}}(\boldsymbol{r}) ~=~  \Gamma
                  \int\limits_{0}^{2\pi}\,\delta(\boldsymbol{r} - \boldsymbol{r}(\xi))\partial_\xi{\boldsymbol{r}}(\xi)d\xi\,.
       \end{equation}

As a consequence of equality (\ref{p_can}) and equality (\ref{vort_w}), the following formula holds:
\begin{equation}
        \label{p_Gamma}
	{\boldsymbol{p}}  ~=~    \pi\varrho_0 {R}^2 \Gamma\, {\boldsymbol b}\,. 
	\end{equation}

As a next step, we redefine the variables $\Delta R$ and  $\phi(\tau)$ as follows:
\[ \chi  ~=~ \frac{\Delta R}{R_f}\cos(\phi_{\,0} +\omega\tau)\,, \qquad  \varpi  ~=~ \frac{\Delta R}{R_f}\sin(\phi_{\,0} +\omega\tau)\,.\]
Clearly, the behavior of the variables $\varpi$ and  $\chi$  is similar to that of a harmonic oscillator.
Finally, we  postulate the set
\begin{equation}
\label{new_set1}
{\cal A}^{\,\prime} ~=~ \bigl\{\, {\boldsymbol p}\,,  {\boldsymbol{q}}\,;\ \varpi\,, \chi\,
  \,\bigr\}
 \end{equation}
as the set of fundamental independent variables for the considered dynamical system -- circle-shaped vortex ring which evolves in accordance with Eq.(\ref{LIE_pert}).

We must note the following important circumstance. The one-to-one correspondence  ${\cal A} \leftrightarrow {\cal A}^{\,\prime} $ takes place  in our (special) case of  round-shaped loops  only. For the perturbed round-shaped loops, as well as the arbitrary closed filaments,  the application  of the formula (\ref{p_can}) for  definition of the new variables leads to the correspondence
\[{\cal A} \leftrightarrow \Omega \subset {\cal A}^{\,\prime}\,,\qquad
{\cal A}^{\,\prime} ~=~ \bigl\{\, {\boldsymbol p}\,,  {\boldsymbol{q}}\,;\ \varpi\,, \chi\,, \dots \scalebox{0.7}{$Kelvin~ waves~ modes$}  \,\bigr\}  \]
where the set $\Omega$ is certain constraint surface. 
This means that passing to the set of variable ${\cal A}^{\,\prime} $, from the set variable ${\cal A} $, is not a trivial operation.  These issues were discussed more detail  in the paper \cite{Tal18}.

Following Dirac's guidance on the fundamental role of the Hamiltonian structure, we axiomatically define it.  The relevant definitions are given below.
   
	\begin{itemize}
  \item Phase space ${\mathcal H} =  {\mathcal H}_{pq}  \times  {\mathcal H}_b   $. The space $ {\mathcal H}_{pq}$ is the phase space for a $3D$  free structureless which     is  parametrized by the variables 
   ${\boldsymbol{q}}$ and  ${\boldsymbol{p}}$.  The space    $ {\mathcal H}_b$  is a phase space for one-dimensional harmonic oscillator.
		 \item Poisson structure:
  \begin{equation}
  \{p_i\,,q_j\}  =  \delta_{ij}\, \quad (i,j = x,y,z)\,, 
  \qquad 
  \{ \varpi, \chi\}  =  \frac{1}{{\cal E}_0 t_0 }\nonumber 
  \end{equation}
  	All other brackets  vanish. 
\item Hamiltonian 
\begin{equation}
        \label{hamilt_1}
				H ~=~ \frac{\boldsymbol{p}^2}{2 \mu_0} ~+~  \frac{{\cal E}_0\,\omega}{2}\Bigl(\varpi^2  + \chi^2  \Bigr)\,.
	\end{equation}
	\end{itemize}
This form of the function $H=H({\boldsymbol p}\,;\ \varpi\,, \chi\,)$ is a direct consequence of our definition of energy -- see Eq.(\ref{E_general})
 Hamiltonian  $H$ provides the dynamics of variables in the {\it conditional time} $t^\# = t_0\tau$.
For example,
\begin{eqnarray}
e^{- t^\# \{H,\dots}\,  \phi(0) &=&   \phi(0) ~+~ (\omega/t_0) t^\#\,, \nonumber \\
e^{- t^\# \{H,\dots}\,  q_i(0) &=& q_i(0) ~-~ ({t^\#}/{\mu_0})p_i\,.\nonumber
\end{eqnarray}

As a consequence of equality (\ref{p_Gamma}), the following formula holds:
\begin{equation}
\label{main_con}
{\boldsymbol p}^{\,2}   ~=~  \pi^2\varrho_0^2 {R_f}^{\,4}\Bigl(1 + \varpi^{\,2} + \chi^{\,2} \Bigr)^{2}\, \Gamma^{\,2}\,.
\end{equation}
This formula shows that the circulation $\Gamma$ is the function of new variables that parametrize the set (\ref{new_set1}).  As previously noted, details can be found in the author's works referenced.

\section{ Circle-shaped quantum vortices \\ as quantized classical systems }

~~~The transformation  ${\cal A} \to {\cal A}^{\,\prime} $  of the classical variables naturally leads to the description of a quantum vortex in terms of the Hilbert space
\begin{equation}
	\label{space_quant}
	\boldsymbol{H}_1  ~=~  \boldsymbol{H}_{pq} \otimes   \boldsymbol{H}_b  \,,
	\end{equation}
			where the symbol   $\boldsymbol{H}_{pq}$  denotes the Hilbert space  of a free structureless particle in the space  $\mathbb{R}_3$.  In this paper we assume
			$\boldsymbol{H}_{pq} = L_2(\mathbb{R}_3)$.  	The	symbol $\boldsymbol{H}_b $ denotes		the Hilbert space  of the quantized harmonic oscillator with classical variables $\chi$ and $\varpi$ (see definitions of these quantities before the formula (\ref{new_set1})).
				This space is formed by the vectors
		\[|\,n\rangle   ~=~  \frac{1}{\sqrt{n!}} (\hat{b}^+)^n    |\,0_b\rangle   \qquad 
		[\,\hat{b}, \hat{b}^+] ~=~ \hat{I}_b\,, \quad \hat{b}|\,0_b\rangle ~=~ 0\,,  \]
			where vector  $|\,0_b\rangle \in \boldsymbol{H}_b$ is vacuum vector and symbols $\hat{b}^+$ and $\hat{b}$ mean the creation and annihilation operators. 	
In accordance with canonical rules, the classical variables  $\varpi$ and $\chi$  become operators
\begin{equation}
	\label{quant_rules}
 \chi ~\to~ \sigma_{ph}\, \frac{\hat{b} + \hat{b}^+}{\sqrt{2}}\,, \qquad 
\varpi ~\to~ \sigma_{ph}\, \frac{\hat{b} - \hat{b}^+}{i\sqrt{2}}\,, \qquad \sigma_{ph} ~=~ \sqrt{\frac{\hbar}{{\mathcal E}_0 t_0}}
\end{equation}

where the dimensionless constant $\sigma_{ph}$
depends on the Planck constant $\hbar$. 
The variables $\boldsymbol q $ and $\boldsymbol p $ are quantized according to the standard rules of quantum mechanics for a free non-relativistic particle in  Hilbert space $L_2(\mathbb{R}_3)$  in a coordinate representation.

At this point, it is appropriate to recall that we are still considering the evolution of our dynamic system in ''conditional time'' $t^\# =  \tau t_0$. In this case, the  vortex ring is simplified to the "3D particle + harmonic oscillator" system.
 As a consequence of Eq.(\ref{hamilt_1}) and the postulated quantization rules, the quantum   Hamiltonian $\widehat{H}^\#$ has the following form:
		\begin{equation}
	\label{H_quant}
		\widehat{H}^\#  ~=~ -~  \frac{\hbar^2}{2\mu_0}\Delta ~+~ \frac{\hbar\, \omega}{t_0}
		\left(b^+ b + \frac{1}{2}\right)\,.
		\end{equation}
	In the final formulas, we will need to return to the real time $t$.  
Spectral problem
		$\widehat{H}^\# |\psi\rangle = E^\#|\psi\rangle$
		has following solutions:
		\begin{equation}
		\label{R_state}
	 \boldsymbol{H}_1^\prime ~\ni~     |\psi_{\boldsymbol{p},n}\rangle ~=~ 
	|\,\boldsymbol{p}\rangle|n\rangle\,, \qquad\quad
		|n\rangle  \in \boldsymbol{H}_b\,,
		\end{equation}
		where  the vector $|\,\boldsymbol{p}\rangle \in  \boldsymbol{H}_{pq}^\prime$  is the eigenvector of the momentum operator $\boldsymbol{\hat p}$.
		Symbol $\boldsymbol{H}_{pq}^\prime$ means corresponding rigged (''equipped'') Hilbert space (see, for example, \cite{BerShu}). 

In earlier writings, the author  outlines  in detail the spectrum of quantized vortex rings. The main steps in determining the spectrum are schematically represented in the diagram:

%%% Diagram	\vspace{5mm}

		{\Large
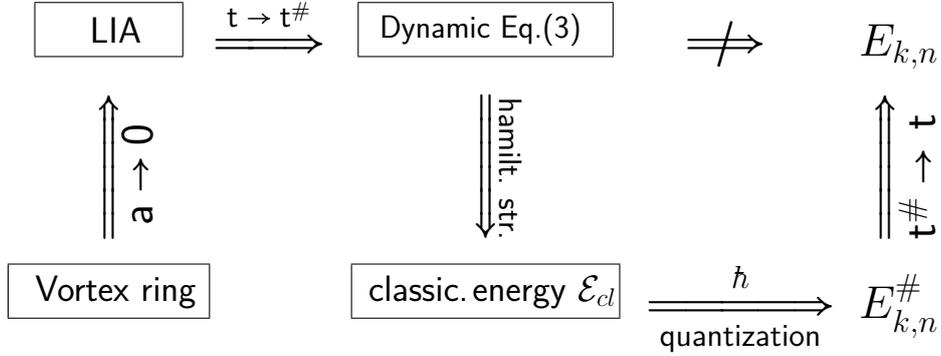
\[
\begin{array}{ccccc}
\framebox(55,20){ \scalebox{0.8}{{\sf LIA}}} &\hspace{-3mm}\xLongrightarrow[\text{~~~}]{\text{${\sf{t}\to \sf{t}^\#}$}} & \framebox(95,20){\scalebox{0.7}{\sf Dynamic Eq.(\ref{LIE_pert}) }} & {\not\Longrightarrow}  & ~~E_{k,n}  \\[3mm]
 \Bigg\Uparrow 
\lefteqn{\begin{turn}{90}${\hspace{-8mm}{~\sf{a} \to 0}}$\end{turn}}
&& \Bigg\Downarrow\lefteqn{\begin{turn}{-90}{\hspace{-12mm}\scalebox{0.7}{ \sf hamilt. str.}}\end{turn} }\vspace{3mm}&&\Bigg\Uparrow
  \lefteqn{\begin{turn}{90}${\hspace{-8mm}{{\sf t}^\# \to\, {\sf t}}}$\end{turn}}\\[5mm]
	\framebox(75,20){ \scalebox{0.8}{{\sf Vortex ring}}} && \framebox(100,20){ \scalebox{0.8}{\sf classic.\!\! energy ${\cal E}_{cl}$}} & \xLongrightarrow[\text{\sf quantization}]{\text{$\hbar$}}\hspace{-4mm}  & ~~E_{k,n}^\# 
    \end{array}
\]}
		
		\vspace{5mm}

Unlike article \cite{Tal_Chaos25J}, we consider here the motion of vortex loops in unbounded space, without any boundary conditions.	
		According to the Eq. (\ref{main_con})  and  postulated quantization rules, we have  the following spectral problem for possible values of circulation $\Gamma$:
			
\begin{equation}
\label{eq_sp_Gamma}
  %      \label{for_Gamma1}
\left[\hbar^2\Delta  ~+~ \pi^2 \varrho_0^2 \Gamma^2 R_f^4  \Bigl({\hat I}  +  \sigma_{ph}^2\,
(\hat{b}^+ \hat{b} + 1/2)  \Bigr)^2\right]|\Psi\rangle ~=~0\,, 
	\end{equation}
where $|\Psi\rangle \in  \boldsymbol{H}_1$. 
Here, as in other places, we assume that the operator $\Delta$ is defined  in  the space $\boldsymbol{H}_1$   as $\Delta \otimes {\hat I}_b $.  The same remark applies to the operators  $\hat{b}$ and      $\hat{b}^+$.  
The eigenvectors  $|\Psi_{\boldsymbol{p},n}\rangle \equiv 
	|\,\boldsymbol{p}\rangle|n\rangle$  of the operator $\widehat{H}^\#$ are also eigenvectors for  the spectral problem (\ref{eq_sp_Gamma}). Corresponding  eigenvalues
	$\Gamma = \Gamma_{k,n}$ will be like this:
	\[ \Gamma_{k,n} ~=~ \epsilon \Gamma_{k,n}^+\,,\]
where 	
\begin{equation}
\label{Gamma_val}
\Gamma_{k,n}^+ ~=~   \frac{\hbar k R_f}{~\tilde\mu_0 \bigl[ 1 +  \sigma_{ph}^2(n + 1/2)\bigr]}\,, \qquad k = |\boldsymbol{k}|, \quad  \boldsymbol{k} = \boldsymbol{p}/\hbar,
\quad n = 0,1,2,\dots.
\end{equation}
and   $\epsilon = {\rm sgn}(\boldsymbol{k}\boldsymbol{b})$ in accordance with Eq.(\ref{p_Gamma}).

When boundary conditions are absent, the $\Gamma$ value's spectrum becomes continuous. This result is expected for a quantized observable in an infinite space. 
For large values of the quantum number $n$ (when $\sigma_{ph}^2n > 1$), the $\Gamma$ value does not depend on the Plank constant $\hbar$ so  this observable becomes quasi-classical.  Let us note that the spectrum of this observable will be discrete if we consider the vortex motion within a certain bounded domain. As established in the previous author's works, the values of the circulation  tends to conventional values $\Gamma \to  \Gamma_m  = \hbar m/\mu_1 $ (number $m$ is integer )  in certain limiting cases. We will not discuss this in detail here.

In the next step, we must go back to "real-time" evolution, and consequently, to the "true" energy spectrum $E_{n}(k)$.This goal can be achieved exactly as described in paper \cite {Tal_Chaos25J}, where the case of the vortex evolution in bounded space domain $D$ was considered. To avoid unnecessary repetition in calculations, we will use some results from this work. 
In the paper \cite{Tal_Chaos25J}  we  considered the spectral problems for the energy and circulation  in a domain $D$, with homogeneous Dirichlet boundary conditions.
 The corresponding   eigenvalues of the Laplace operator $\Delta$ were denoted   as $- (\lambda_{[m]}/L)^2$.   Therefore, we can use the deduced formula for the  energy $E_{[m],n}$ from the paper  \cite{Tal_Chaos25J} by making the  replacement   $ \lambda_{[m]} \to k L$.
The result is written as follows:

\begin{equation}
	\label{energy}
 E_{n}(k) ~=~ \frac{\hbar^2 k}{8\pi\tilde\mu_0 R_f}  
 \frac{\bigl[\omega\, (2n + {1}) ~+~ {\sigma_{ph}^2}(k L)^2 \bigr]}{\bigl[1 +  \sigma_{ph}^2(n + 1/2)  \bigr]^2}
\,,
\end{equation}

The second term in the square brackets of  numerator  is proportional to  the $\hbar$ value.   Therefore, the  summand $\sigma_{ph}^2(k L)^2$
  becomes significant  if the momentum  $\boldsymbol{p} = \hbar\boldsymbol{k}$ 
	becomes   very large.  Moreover, only the second term  in the square brackets  of formula (\ref{energy})  depends on the scale factor $L$. 	Thus, this summand can be considered as a 
''scale correction'' for the energy value.

Finally, we can write a formula for the Hamiltonian which provides the real-time evolution of the vortex rings under consideration:
\begin{equation}
	\label{Hamilt_fin}
 {\widehat H}(\omega) ~=~ \sum_{n} \int d^{\,3} \boldsymbol{p}\, E_{n}(k)\,|\Psi_{\boldsymbol{p},n}\rangle\langle\Psi_{\boldsymbol{p},n}|\,. 
  \end{equation}
The state space of the  quantum system with  $K$ non-interacting  circle-shaped vortices  is as follows:
\begin{equation}
	\label{H_N}
\boldsymbol{H}_K ~=~  \underbrace{\,\boldsymbol{H}_1 \otimes\dots\otimes\boldsymbol{H}_1\,}_{K}\,.
\end{equation}

\section{Quantum vortex loops of arbitrary size and  shape }

~~~In a certain sense, our method for the   vortex rings quantization  uses the fact that the solution (\ref{our_sol}) conserves its shape in the dynamics.  Formula (\ref{p_can}) yields the simple equation (\ref{main_con}) only under these conditions.
Of course, we can consider small-ring oscillations (Kelvin waves) and then quantize them using the proposed method \cite{Tal22_1}.  However, our subsequent goal is to describe the quantum vortex loops with an arbitrary shape and size. 
To commence our construction, we begin our research by examining finite closed filaments.  Finite vortices that have a beginning and an end at the boundaries of the domain are beyond the scope of our model. We also do not consider infinite vortex filaments that are apparently not possible in practice.

 To achieve the goal declared, within the framework of our approach, it is necessary to make some additional assumptions.
At the beginning, we introduce the following notation.
\begin{itemize}
\item $\boldsymbol{r}({\sf s})$ -- radius-vector of a smooth closed curve (''vortex filament'') that was parametrized by natural parameter ${\sf s}$;
\item $R({\sf s})$ --  radius of curvature    of the curve $\boldsymbol{r}({\sf s})$ at a point ${\sf s}$:
\begin{equation}
\label{Rs}
     \frac{1}{R({\sf s})} ~=~ \left\vert\frac{d^2 \boldsymbol{r}({\sf s})}{d{\sf s}^2 }\right\vert\,.
\end{equation}   
 To simplify our studies, we assume that curvature $1/R({\sf s}) \not= 0$ along the curve $\boldsymbol{r}({\sf s})$.  We will return to considering flex-points later;
\item $\boldsymbol{q}({\sf s})$ -- evolute of the curve  $\boldsymbol{r}({\sf s})$:
\begin{equation}
\label{evolute}
 \boldsymbol{q}({\sf s}) ~=~ \boldsymbol{r}({\sf s}) ~+~  R^{\,2}({\sf s}) \frac{d^2 \boldsymbol{r}({\sf s})}{d{\sf s}^2 }\,.
\end{equation}
In accordance with our assumptions, the curve $\boldsymbol{q}({\sf s})$ is a compact closed curve with a cuspidal points ${\sf s}_0,\dots,{\sf s}_m$ that a correspond to the extremal points of the periodic function $R = R({\sf s})$.
Because the parameter ${\sf s}$ is not  a natural parameter for the curve $\boldsymbol{q}({\sf s})$, it is convenient to introduce a natural parameter here. For any smooth part of the curve $\boldsymbol{q}({\sf s})$, this parameter is defined as:
\begin{equation}
\label{ls_funct}
\ell ~\equiv~ \ell({\sf s}) ~=~ \sum_{i=0}^n I_i ~+~ \int_{{\sf s}_n}^{{\sf s}}\sqrt{\boldsymbol{q}^\prime({\sf s})}\, d{\sf s}\,, \qquad {\sf s} \in [{\sf s}_n, {\sf s}_{n+1})\,,
\end{equation}
where
\[ I_0 ~=~ 0\,, \qquad  I_n ~=~ \int_{{\sf s}_{n-1}}^{{\sf s}_n}\sqrt{\boldsymbol{q}^\prime({\sf s})}\, d{\sf s}\,, \quad n = 1,2,\dots\]
\item $|\beta\rangle$ -- coherent state  \cite{Perelomov}  for the harmonic oscillator (the eigenvector for the annihilation operator $\hat b$):
\[ \hat b \,|\beta\rangle ~=~ \beta |\beta\rangle\,, \qquad  |\beta\rangle \in \boldsymbol{H}_b  .\] 
\end{itemize}

In accordance with our quantization rules, the radius of a circular  vortex loop becomes  the operator as
\[ R^{\,2} ~\longrightarrow~ \hat{R}^{\,2} ~=~ R_f^{\,2}
\left[{\hat I}_b + \sigma_{ph}^2\left({\hat b}^+ {\hat b} + \frac{1}{2}\right)\right]\,.\]   
This operator  has the following eigenvalues  ${R}_n$:
\begin{equation}
\label{spectrum_R}
 {R}_n ~=~ R_f\sqrt{1 + \sigma_{ph}^2\left(n + \frac{1}{2}\right)} \,,\nonumber
\qquad  n ~=~ 0,1,\dots, \,.
\end{equation}
Let us consider the tangent circle for the curve $\boldsymbol{r}({\sf s})$ in a point ${\sf s}$.  We correspond the coherent state $|\beta\rangle = |\beta({\sf s})\rangle$  to the radius $R({\sf s})$ by the formula:
\begin{equation}
\label{Rs_beta}
R^{\,2}({\sf s}) ~\equiv~  \langle\beta({\sf s}) |\hat{R}^{\,2}|\beta({\sf s})\rangle ~=~
R_f^{\,2}
\left[1 + \sigma_{ph}^2\left(|\beta({\sf s})|^{\,2} + \frac{1}{2}\right)\right]\,.
\end{equation}
This formula defines the value $|\beta({\sf s})|$ from the radius $R({\sf s})$ uniquely.
As regards the argument $\arg\beta({\sf s})$, we assume that  this value does not depend on the curve parameter ${\sf s}$:  $\arg\beta({\sf s}) \equiv const$. 
In accordance with our quantization postulates  (\ref{quant_rules}), the correspondence
\[ \frac{1}{\sqrt{2}}\Bigl(\chi + i \varpi\Bigr) ~\longrightarrow~ \sigma_{ph} \hat b\,\]
holds. That is why the following equality for the function $\arg \beta({\sf s})$ takes place:
$\arg \beta({\sf s}) ~=~ \phi_0 ~+~ \omega\tau\,.$ Therefore, we have the following statement for the flow  $\Phi_\omega$ along the curve  $\boldsymbol r(s)$:
\begin{equation}
\label{flow1}
\Phi_\omega({\sf s}) ~\propto~ \omega {\sf a}^2\,, \qquad  \frac{d \Phi_\omega({\sf s})}{d {\sf s}}  ~\equiv~ 0\,.
\end{equation}

In  the article \cite{Tal_PoF23} author explores the initial phase of turbulence through the lens of their proposed method. This phase is characterized by the creation and subsequent division of individual circular micro-vortices.
In this regard, it seems appropriate to ask the following question: how do large vortex filaments form, which eventually lead to the formation of a vortex tangle?
Next, we will introduce a quantum state which, in our opinion, that can be interpreted as
 a ''large'' vortex filament $\boldsymbol{r}({\sf s})$. To make this concept clearer, we will explore it through three stages.

{\bf 1.} Let us  consider the entangled   state 
$|\,\Phi[\boldsymbol{r}({\sf s})]\rangle \in \boldsymbol{H}_1^\prime$:
\begin{equation}
\label{Phi_q}
|\,\Phi[\boldsymbol{r}({\sf s})]\rangle ~=~  \oint f({\sf s})|\boldsymbol{q}({\sf s})\rangle |\beta({\sf s})\rangle \,d {\sf s}\,,   \qquad  \oint |f({\sf s})|^2d {\sf s} = 1\,.
\end{equation}
In this definition,  both the function $\boldsymbol{q}({\sf s})$ and the vector-function $|\beta({\sf s})\rangle$ are defined,  as it was described above.
It is clear that the vector (\ref{Phi_q}) describes the continuous coherent superposition of the vectors $|\boldsymbol{q}({\sf s})\rangle |\beta({\sf s})\rangle$ that correspond to a circular vortex filament of radius $R({\sf s})$  (see Eq.(\ref{Rs_beta})) and  position $\boldsymbol{q}({\sf s})$ of the circle center. 
The curve $\boldsymbol{r}({\sf s})$ is envelope of a family of corresponding circles.
Because our assumption $\arg\beta({\sf s}) \equiv const$, the value $\arg\beta$ defines certain flow in a curve $\boldsymbol{r}({\sf s})$.

Note that this interpretation of a quantum vortex line $\boldsymbol{r}({\sf s})$  inherently includes anisotropy relative to the tangent vector $d\boldsymbol{r}({\sf s})/ds$  from the start. We will not develop this topic any further here.
Let us only note that the anisotropy in a vortex tangle has been discussed early within the framework of the two-fluid model for super-fluidity \cite{Nemir2}. Among other things, this work noted that the problem of the vortex tangles anisotropy has not yet been theoretically explained.

 {\bf 2.}   Although representation (\ref{Phi_q}) is quite visually, it is convenient to use a different representation of the considered filament-like object.
The reason is that the eigenvectors $|\boldsymbol{q}\rangle$ of the position operator
$\hat{\boldsymbol{q}}$ correspond to the unphysical states, because $|\boldsymbol{q}\rangle \not\in \boldsymbol{H}_{pq} =  L_2(\mathbb{R}_3)$.
 To avoid unphysical states,   we will use coherent states
$|\boldsymbol{z}\rangle  \in \boldsymbol{H}_{pq} $ for  free  particles \cite{BagGit} instead of the vectors  $|\boldsymbol{q}\rangle \in \boldsymbol{H}_{pq}^\prime$. In  paper \cite{BagGit}, the time-dependent coherent states  $|{z},t\rangle$   of a free non-relativistic  particle in one space dimension were considered.  We are considering short-lived vortices\footnote{In the future, our main goal is to describe the vortex tangle of a turbulent flow  in detail.  The lifetime of vortices in a turbulent flow is small; this value  restricted by the period 			$T = 2\pi/W$, where symbol $W$ means the vorticity  \cite{Tenn}. }, so we will use coherent states  $|{z}\rangle  \equiv |{z},0\rangle$ only. 
We will not reproduce the definition and the process of constructing coherent states $|{z}\rangle$   here.  The result is the following expression for the particle wave function in a $\boldsymbol{q}$-representation: 

\begin{equation}
\label{Psi_q}
\Psi(\boldsymbol{q}) ~\equiv~   \langle \boldsymbol{q}|\boldsymbol{z},0\rangle ~=~
{C_\varepsilon} \exp\left[\frac{i}{\hbar}\boldsymbol{p}_0\boldsymbol{q} - \frac{1}{4\varepsilon^2}
\Bigl(\boldsymbol{q} ~-~ \boldsymbol{q}_0\Bigr)^2    \right]\,,
\end{equation}
where  $\boldsymbol{p}_0 = \boldsymbol{p}(t)\vert_{t=0}$,  $\boldsymbol{q}_0 = \boldsymbol{q}(t)\vert_{t=0}$  and symbol $C_\varepsilon$ stands for the normalizing constant.   To generalize the results of the paper \cite{BagGit} on a $3D$ case, we assume that $\varepsilon_x  =  \varepsilon_y  =  \varepsilon_z  =  \varepsilon$.

Let us return to discussed vortex filament ${\boldsymbol r}({\sf s})$.
Firstly, we pass from  $\boldsymbol{q}$-representation in  Eq.(\ref{Psi_q}) to 
$\boldsymbol{p}$-representation.  Secondly, we make replacement
\[ \boldsymbol{q}_0 ~\longrightarrow~ \boldsymbol{q}({\sf s})\,, \qquad
\boldsymbol{p}_0 ~\longrightarrow~  {\sf p}(\ell) \boldsymbol{\rm b}({\sf s}) ~\equiv~
\hbar {\sf k}(\ell)\boldsymbol{\rm b}({\sf s}) \,,\qquad
\ell = \ell({\sf s})\,,\]
where evolute $\boldsymbol{q}({\sf s})$ was defined by Eq.(\ref{evolute}), function
$\ell({\sf s})$ was defined by Eq.(\ref{ls_funct}), vector 
$\boldsymbol{\rm b}({\sf s})$ means binormal unit vector for the curve ${\boldsymbol r}
({\sf s})$ and the function ${\sf p}(\ell) \equiv \hbar {\sf k}(\ell)$ some smooth periodical function defined on the curve $\boldsymbol{q}({\sf s}) \equiv  \boldsymbol{q}(\ell({\sf s}))$.  The outcomes will be as follows:
\begin{eqnarray}
\label{WaveF_p}
~&~&~\Psi(\boldsymbol{p}; {\sf s})  ~\equiv~   \langle \boldsymbol{p}|\boldsymbol{z}({\sf s})\rangle ~=~\\[2mm]
~&=&~B_\varepsilon \exp\left[- \frac{i}{\hbar} \Bigl(\boldsymbol{p}  - {\sf p}\bigl(\ell({\sf s})\bigr) \boldsymbol{\rm b}({\sf s})\Bigr) \boldsymbol{q}({\sf s}) 
  - \frac{\varepsilon^2}{\hbar^2}
\Bigl(\boldsymbol{p} ~-~ {\sf p}\bigl(\ell({\sf s})\bigr) \boldsymbol{\rm b}({\sf s})\Bigr)^2    \right]\,,\nonumber
\end{eqnarray}
where the normalizing constant $B_\varepsilon$  is as follows:
\[ B_\varepsilon ~=~  \left(\frac{\varepsilon}{\hbar}\right)^{3/2} 
\left(\frac{2}{\pi}\right)^{3/4}\,.\]

 Let us define the  entangled  state $|\,\Psi_{\boldsymbol{r},p}\,\rangle \in \boldsymbol{H}_1$:

\begin{equation}
\label{Psi_line}
|\,\Psi_{\boldsymbol{r},p}\,\rangle ~\equiv~
|\,\Psi[\boldsymbol{r}({\sf s}); {\sf p}(\ell)]\,\rangle ~=~  \oint f({\sf s})|\,\boldsymbol{z}({\sf s})\rangle |\,\beta({\sf s})\rangle \,d {\sf s}\,.
\end{equation}
In accordance with the arguments above, we declare that quantum state (\ref{Psi_line}) corresponds with vortex  filament-like  object ${\boldsymbol r}({\sf s})$.
%%%%%%%

{  Now, let us discuss the mechanism of how entangled quantum states  (\ref{Psi_line})  appear. 
The initial stage of a quantum  turbulent flow  was studied in the author's  previous  paper  \cite{Tal_PoF23}. In accordance with the suggested mechanism, the appearance of a single circular micro-vortex due to fluctuation (for example) leads to subsequent divisions of micro-vortices. Let a single  micro-vortex is described by the vector
\[ |\,\psi_{g;n}\,\rangle ~=~ \int g(\boldsymbol{q})|\, \boldsymbol{q} \,\rangle |\,n\,\rangle
~\in~ \boldsymbol{H}_1\,, \qquad   g(\boldsymbol{q}) \in L_2(\mathbb{R}_3)  \,.\]
It is easy to see that the amplitude $\langle\,\psi_{g,n}\,   |\,\Psi_{\boldsymbol{r},p}\,\rangle$   will generally be non-zero:

\begin{equation}
\label{amplitude}
  \langle\,\psi_{g,n}\,   |\,\Psi_{\boldsymbol{r},p}\,\rangle ~=~
	\int d\boldsymbol{q}  \oint d {\sf s} g(\boldsymbol{q})  f({\sf s})\langle\,\boldsymbol{q}\, |\,\boldsymbol{z}({\sf s})\rangle \langle\, n\, |\,\beta({\sf s})\rangle \,d {\sf s}\, ~\not=~ 0\,.
\end{equation}
Indeed, the function $\langle\,\boldsymbol{q}\, |\,\boldsymbol{z}({\sf s})\rangle $ was presented  above (see Eq.(\ref{Psi_q}) and the function $\langle\, n\, |\,\beta({\sf s})\rangle = \beta({\sf s})^n\exp(-|\beta({\sf s})|^2/2)/\sqrt{n!}$, where the value $|\beta({\sf s})|$ was defined by the  Eq.(\ref{Rs_beta}).
In accordance with basic  quantum theory principles, the value  $|\langle\,\psi_{g,n}\,   |\,\Psi_{\boldsymbol{r},p}\,\rangle|^2$ gives the probability for the 
pass from the state $|\psi_{g,n}\rangle$  to  the constructed entangled state $|\,\Psi_{\boldsymbol{r},p}\,\rangle$.

Following the  developed  model \cite{Tal_PoF23}, 
  we have a number $N$ of micro-vortices after some time has passed since the appearance of the first vortex loop. The corresponding state of the quantum system is a vector of space $\boldsymbol H_N$ (see Eq.(\ref{H_N})). This fact means that the probability   of an entangled state (\ref{Psi_line}) appearing  in any space $\boldsymbol H_1$ becomes   
$N|\langle\,\psi_{g,n}\,   |\,\Psi_{\boldsymbol{r},p}\,\rangle|^2$.
For sufficiently large values of $N$, the probability tends to one for every curve
$\boldsymbol{r}({\sf s})$.  Thus,  at different times, an entangled state  (\ref{Psi_line})  appears for every random  functions $f({\sf s})$ and $\boldsymbol{r}({\sf s})$ that satisfy the condition (\ref{amplitude}).}

{\bf 3.} The pure state (\ref{Psi_line})  corresponds to a coherent superposition  of vectors   $|\,\boldsymbol{z}({\sf s})\rangle |\,\beta({\sf s})\rangle$.  We can use the operator $\hat\rho_0   = |\,\Psi_{\boldsymbol{r},p}\,\rangle\langle\,
\Psi_{\boldsymbol{r},p}\,|$  (''density matrix'') instead of the vector $|\,\Psi_{\boldsymbol{r},p}\,\rangle$,  It is well-known that this replacement be equivalent in a coherent case. Obviously, the corresponding coefficient matrix $\overline{f({\sf s})}f({\sf s}^\prime)$ in the decomposition of  operator $\hat\rho_0$  has  off-diagonal elements. 
For our subsequent studies, we intend to consider a vortex filament as an element of a vortex tangle. Corresponding interactions lead to  decoherence:
\[ \overline{f({\sf s})}f({\sf s}^\prime) ~\longrightarrow~ |f({\sf s})|^2 \delta({\sf s} - {\sf s}^\prime)\,.\]
This means that the reduction $\hat\rho_0 \to \hat\rho$ occurs, 
where the final density matrix  $\hat\rho \equiv \hat\rho[\boldsymbol{r},{\sf p}]$  has the  following form:
\begin{equation}
\label{rho_final}
\hat\rho ~=~  \oint |f({\sf s})|^2  |\,\boldsymbol{z}({\sf s})\rangle |\,\beta({\sf s})\rangle \langle\, \beta({\sf s})|\langle\,\boldsymbol{z}({\sf s})|  \,d {\sf s}\,.
\end{equation}

Our studies demonstrate that the continuous entangled superposition of the states
 $|\,\boldsymbol{z}(s)\rangle |\, \beta(s)\rangle$ 
forms a specific  filament ${\boldsymbol r}({\sf s})$
which is connected to  constant  internal  flow (\ref{flow1}).
  We interpret such superposition as a creation of a ''large'' vortex filament. 
	Taking into account the surrounding media, we must consider a statistical mixture (\ref {rho_final}), instead of the initial pure quantum state.

$\star$ Finally, we declare that the density matrix (\ref{rho_final}) corresponds to the quantum state of our vortices-formed curve ${\boldsymbol r}({\sf s})$.  We interpret it as a     ''large''  vortex (or vortex-like)  loop.

Taking into account Eq.(\ref{Hamilt_fin}), we can deduce the following formula for energy of a filament ${\boldsymbol r}({\sf s})$:
\begin{equation}
\label{Energy_r}
E[{\boldsymbol r}({\sf s})] ~=~ Tr\Bigl( \hat\rho \hat H \Bigr) ~=~ \oint |f({\sf s})|^2 
E({\sf s})\,d {\sf s}\,
\end{equation}
where value $E({\sf s})$ is the   vortex energy in the  state $|\,\boldsymbol{z}({\sf s})\rangle |\,\beta({\sf s})\rangle$:
\begin{eqnarray}
\label{E_s}
 E({\sf s}) &=& \langle\, \beta({\sf s})|\langle\,\boldsymbol{z}({\sf s})| \hat H|\,\boldsymbol{z}({\sf s})\rangle |\,\beta({\sf s})\rangle ~=~
|B_\varepsilon|^2 \exp\Bigl[ - \,|\beta({\sf s})|^2\, \Bigr] \times\\[2mm]
~&\times& \!\! \sum_{n=0}^\infty\frac{\,|\beta({\sf s})|^{2n}}{n!}\int E_n(k) \exp\left[ - {2 \varepsilon^2}\Bigl( \boldsymbol{k} ~-~ {\sf k}\bigl(\ell({\sf s})\bigr) \boldsymbol{\rm b}({\sf s})  \Bigr)^2 \,\right]\,d^{\,3} \boldsymbol{p}\,.\nonumber
\end{eqnarray}

In this formula, the  function $E_n(k)$ is defined by Eq.(\ref{energy}) and, as usual, $\boldsymbol{p} = \hbar\boldsymbol{k}$.
It is clear that both the series and the integral converge in Eq.(\ref{E_s}).

\section{The mechanism of disconnection  of a large loops.}

~~~ { One of the main differences between the proposed approach and traditional ones is that it predicts a richer spectrum of circulation.}
Conventional  quantum vortex theory considers a vortex filament as a topological defect, where the circulation $\Gamma = (\hbar/\mu_1) n$, $n = 1,2,\dots$ remains consistent throughout the entire length of the filament.
 
 The arguments in favor of the extending   this  spectrum, as well as a more detailed comparison with other approaches, were presented in the author's previous works (see, for example, \cite{Tal_PoF23,Tal_PhScr}) To avoid repetition, we will focus here only on the value of vortex circulation in turbulent flow.  Nevertheless, we briefly point out the difference between our approach and the standard one:
%\vspace{5mm}

{\small
\begin{tabular}{|l||l|l|} \hline
{\sf ~~~Considered object} &{\sf Conventional approach }& {\sf ~~~Our approach} \\ \hline\hline
Vortex energy, definition  & By means of   & With help of Cazimir \\
for the thin filaments    & hydrodynamical    & functions of a central   \\
	   in the classical theory  &     calculations        &  extended  Galilei group  \\ \hline
Accounting for the move-&Fluid velocity $\boldsymbol{u}(\boldsymbol{r})$&Circulation as additional\\ 
		ment of surrounding fluid &  & dynamical variable\\ \hline
 Circular quantum  & Topological defect  &  Quantized classical \\
 micro- and meso-vortices  & (wave function$^*$ zeros)    & dynamical system \\ \hline
Quantum vortex as  & Topological defect  &  Entangled state of meso- \\
a large size filament & (wave function$^*$ zeros) &   and micro-vortices \\ \hline
Circulation, kind of  & Topological nature & Eigenvalue of the  \\ 
of the quantum number &  ~~~~~~            & spectral  problem  \\ \hline
Circulation, values in  & ~~~~~~~$\Gamma \propto \hbar n,$ & $\Gamma = \Gamma_{m,n,\dots}\,,$ discrete\\         
the bounded domain  &  ~~~ $(n = 0,1,2,\dots)$    & numbers depending on\\
  $D_3 \subset \mathbb{R}_3$ &    ~~~~~                     &   the kind of domain $D_3$ \\ \hline			
Circulation, values in & ~~~~~~~$\Gamma \propto \hbar n,$ & $\Gamma = \Gamma(p)$, takes \\
 unbounded space $\mathbb{R}_3$ & ~~~ $(n = 0,1,2,\dots)$  & continuous values \\ \hline
Separate fragments  &Fact of their      & Fragments existence   \\
of the vortex filaments & existence is unclear   & follows  from the theory \\ \hline
Quantum vortex   &  Mostly quality    &  Formalizm of quantum \\
interaction      &  scenario          &  many-body theory \\ \hline
\end{tabular}}

\vspace{2mm}

$^*$ In the two-fluid model.

\vspace{5mm}

The known experimental measurements of the $\Gamma$ value  were made for separate isolated  vortices.   An overview of the results on this issue is given in the book \cite{Donn}. 
In the present day,   the formula $\Gamma \propto n$  was confirmed by the results of numerical modeling  in the framework of the Gross-Pitaevskii model \cite{MuPoKr}. 
  These works predict the ''large'' peaks when  $\mu_1\Gamma /\hbar =  n$ for integer numbers $n$ and certain ''small'' peaks when  $\mu_1\Gamma /\hbar \not=  n$. Although these "superfluous" (non-integer) values  were explained by errors, alternative models are quite admitted for such a complex phenomenon as quantum turbulence. Why do we believe this? 

So, the conventional formula for the circulation  means that there is a strong correlation between any points of the fluid at large distances near the vortex core.
Indeed, taking into account the definition of the circulation, we can estimate the fluid velocity $\boldsymbol u(\boldsymbol r)$ near the vortex core:
\[\Gamma ~=~ \frac{\hbar}{\mu_1} n ~\sim~ 2\pi{\sf a}|\boldsymbol{u}(\boldsymbol{r}_1)|~=~ 2\pi{\sf a}|\boldsymbol{u}(\boldsymbol{r}_2)|  \,.\]
Thus, we have a strong correlation between the fluid velocities even over large distances
$|\boldsymbol{r}_1 - \boldsymbol{r}_2|$.   It is quite natural for a single vortex filament in a quantum fluid.  
As it seems, the existence of this correlation in a turbulent flow looks amazing.

Let us return to our model and define  the operator $\hat\Gamma^+$:
\begin{equation}
\label{Gamma_plus}
\hat\Gamma^+ ~=~ \sum_{n=0}^\infty \int d^{\,3} \boldsymbol{p}\, \Gamma_{k,n}^+\,|\,\Psi_{\boldsymbol{p},n}\,\rangle\langle\,\Psi_{\boldsymbol{p},n}\,|\,, 
  \end{equation}
where positive real numbers $\Gamma_{k,n}^+$ were defined by Eq.(\ref{Gamma_val}).
It is easy to  state that the series (\ref{Gamma_plus})  converges in the  strong operator topology (SOT).
It is clear that these numbers are eigenvalues of operator $\hat\Gamma^+$ with eigenvectors $|\,\Psi_{\boldsymbol{p},n}\,\rangle$.

The expectation value of the ''observable'' $\hat\Gamma^+$ for the filament  ${\boldsymbol r}({\sf s})$  is as follows:
\begin{equation}
\label{Gamma_plus_v}
\Gamma^+[{\boldsymbol r}({\sf s})] ~=~ Tr\Bigl( \hat\rho\, \hat\Gamma^+ \Bigr) ~=~ \oint |f({\sf s})|^2 \Gamma^+({\sf s})\,d {\sf s}\,.
\end{equation}
The function $\Gamma^+({\sf s})$ is calculated as
\begin{eqnarray}
\Gamma^+({\sf s}) &=&  \frac{|B_\varepsilon|^2 R_f}{\tilde\mu_0} \exp\Bigl[ - \,|\beta({\sf s})|^2\, \Bigr]  \sum_{n=0}^\infty\frac{\,|\beta({\sf s})|^{2n}}{n![1 +  \sigma_{ph}^2(n +1/2)]}  ~\times \nonumber \\[2mm]
\label{Gamma_s}
~&\times&\!\! \int p\, \exp\left[ - {2\, \varepsilon^2}\Bigl( \boldsymbol{k} ~-~ {\sf k}\bigl(\ell({\sf s})\bigr) \boldsymbol{\rm b}({\sf s})  \Bigr)^2 \right]\,d^{\,3} \boldsymbol{p}\,.
\end{eqnarray}
In accordance with Eq.(\ref{p_Gamma}),
the local  circulation value for some point    ${\sf s}$  
is calculated  as follows:
\[ \Gamma({\sf s}) ~=~  {\rm sgn}\bigl[{\sf k}\bigl(\ell({\sf s})\bigr)\bigr]\, \Gamma^+({\sf s})\,.\]
Thus, the circulation $\Gamma$ does not have a certain constant value along considered curve
${\boldsymbol r}({\sf s})$\,!
Of course, it is impossible for the classical vortex filaments. But in a quantum case the super-fluidity effects can arise.  
This means, in particular, that there is no friction between adjacent layers of fluid.
Therefore, there are no reasons to forbid the different meanings of vorticity
 ${\boldsymbol w}_\parallel({\sf s})$  for the different points  ${\sf s}$ of the filament ${\boldsymbol r}({\sf s})$.
 { Thus, the dependence
\begin{equation}
        \label{vort_ws}
     {\boldsymbol w}_{||}({\sf s}) ~=~  \Gamma({\sf s})
                  \oint\delta(\boldsymbol{r} - \boldsymbol{r}({\sf s}))\partial_{\sf s}{\boldsymbol{r}}({\sf s})d{\sf s}\,
       \end{equation}     
takes place for the considered vortex-like object  ${\boldsymbol r}({\sf s})$.}
Consequently, we have the important corollary   here. Let symbol $\gamma$ means closed curve that does not link both  considered curve  ${\boldsymbol r}({\sf s})$ and circle vortices that correspond to states $|\boldsymbol{q}({\sf s})\rangle |\beta({\sf s})\rangle$.
{ Then, 
our conclusion about the behavior of the  $\Gamma({\sf s})$  means     that inequality 
\begin{equation}
\label{Second_1}
\oint\limits_\gamma  \boldsymbol u({\boldsymbol r}) d {\boldsymbol r} ~\not=~ 0
\end{equation}
holds for the surrounding fluid velocity   $\boldsymbol u({\boldsymbol r})$! 
  Indeed, let us consider some point  ${\boldsymbol r}_0 \equiv {\boldsymbol r}({\sf s}_0)$ on the filament  ${\boldsymbol r}({\sf s})$ and some small $\epsilon$-neighborhood
 ${\cal U}_\epsilon({\boldsymbol r}_0)$   of this point. 
 In accordace with our construction, the core radius ${\sf a}$  is a constant value along the curve ${\boldsymbol r}({\sf s})$ (see Eq.(\ref{flow1}) and  the discussion above).
The neighborhood  ${\cal U}_\epsilon({\boldsymbol r}_0)$ cuts some ''cylinder'' from the filament's  core.
Let ${\boldsymbol r}({\sf s}) \in {\cal U}_\epsilon({\boldsymbol r}_0)$ and the symbol $S_a$ denotes a circle that is formed   by the  cross-section of  this ''cylinder'' by the transversal  plane at  the point ${\boldsymbol r}({\sf s})$. We also denote the  fluid velocity that is tangent to the circle $S_a$  as $\boldsymbol{u}_t({\sf s})$.   
Then, the following dependence $u_t = u_t({\sf s})$   is  true for the neighborhood ${\cal U}_\epsilon({\boldsymbol r}_0)$: 
\[ u_t({\sf s}) ~\equiv~ |\boldsymbol{u}_t({\sf s})|   ~\approx~ u_t({\sf s}_0) ~+~
\frac{({\sf s}-{\sf s}_0) }{2\pi {\sf a}}\left(\frac{\Gamma({\sf s})}{d {\sf s}}\right)\bigg\vert_{\sf s = s_0}  ~+~  {\cal O}\bigl(({\sf s}-{\sf s}_0)^2\bigr)  \,.  \]
Consequently, the absolute value $w_\perp$ of the vorticity ${\boldsymbol w}_\perp$   is defined by the  equation
\begin{equation}
\label{vort_perp}
   w_\perp  ~\equiv~ |\,{\rm rot}\, u({\sf s})| ~=~ \frac{1}{2\pi {\sf a}}\left(\frac{\Gamma({\sf s})}{d {\sf s}}\right) ~\not=~ 0  
\end{equation}
on the each point on the  cylinder surface  and the nearby points. 
As for the direction of the vorticity vector  ${\boldsymbol w}_\perp$, the following relationship holds:  
\[{\boldsymbol w}_\perp \partial_s{\boldsymbol r}({\sf s}) ~=~ 0\,.\] }
 This equality demonstrates that each secondary vortex is perpendicular to  the 
line $\boldsymbol r({\sf s})$ --  filament-like vortex object    under consideration.  There are no other restrictions on the direction of vector ${\boldsymbol w}_\perp$.

%%%%%%%%%%%%%%%%%%%%%%%%%%%%%%%%%%%%%%%%%%%%%%%%%%%%%%%%%%%
 Therefore, summarizing the above considerations,we conclude that   the creation a considered quantum system (an entangled state with a continuous superposition 
 of states that correspond to the isolated 
    circle-shaped micro-vortices)   leads to both a  large filament-like  vortex structure  and the appearance of "secondary" vortices. 
 These additional vortices are not isolated.  The vector field ${\boldsymbol w}_\perp $ is continuous in a neighborhood of the curve ${\boldsymbol r}({\sf s})$. 
 In a sense, secondary vortices can be considered as  point-like vortices distributed continuously near the core of a filament ${\bf r}({s})$.
The density of these vortices near the line ${\boldsymbol r}({\sf s})$ is proportional to the value ${\Gamma({\sf s})}/{d {\sf s}}$: indeed, in accordance with Eq.(\ref{vort_perp}), the vorticity $w_\perp = 0$ if ${\Gamma({\sf s})}/{d {\sf s}}=0$. As for the thickness of the secondary vortex layer, it can approximately be considered equal to radius $R({\sf s})$.  This radius is related to the primary micro-vortices that form the filament ${\boldsymbol r}({\sf s})$  and is determined by Eq.(\ref{Rs}). Probably, the process of separating these vortices from the core is possible. Details will depend on the environment around our object ${\boldsymbol r}({\sf s})$ and require further study. 
  As other vortex structures, these  are short-lived objects in a turbulent flow \cite{Tenn}. 

Next, we will investigate the behavior of the circulation $\Gamma(s)$ for the case
 $R({\sf s}) \to\infty$. At the beginning, we note that the inequality 
\[\frac{1}{1 +  \sigma_{ph}^2(n +1/2)}  ~<~  \frac{2}{\sigma_{ph}^2(2n + 1)}\]
holds. This fact leads to a consequence for the $|\beta(s)|$-depended factor in the first line of Eq.(\ref{Gamma_s}):
\[\exp\Bigl[ - \,|\beta({\sf s})|^2\, \Bigr]  \sum_{n=0}^\infty\frac{\,|\beta({\sf s})|^{2n}}{n![1 +  \sigma_{ph}^2(n +1/2)]} ~<~ \frac{2}{\,\sigma_{ph}^2|\beta(s)|\,}D_+\bigl(|\beta({\sf s})|\bigr)\,,\]
where $D_+(x)$ means the Dauson function:
\[  D_+(x) ~=~  e^{-x^2}\int_0^x e^{t^2}dt\,.\]
Taking into account the asymptotic behavior $D_+(x)\to 1/(2 x)$ for $x \to \infty$ we deduce:
\[ \Gamma({\sf s}) ~\longrightarrow~  const\cdot |\beta({\sf s})|^{\,-2}\,, \qquad\quad
|\beta(s)| \to\infty\,.\]
Finally, because Eq.(\ref{Rs_beta}), we deduce following asymptotic behavior:
\begin{equation}
\label{asym_Gamma}
\Gamma({\sf s}) ~\longrightarrow~  const\cdot |R({\sf s})|^{\,-2}\,, \qquad\quad
R({\sf s}) \to\infty\,.
\end{equation}

To simplify our initial constructions, we assumed the absence of any flex points on the filament ${\boldsymbol r}({\sf s})$.  Of course, this assumption may be true for a fixed time moment, but it is not justified in terms of dynamics. In general, flex points appear and disappear in any vortex filament, especially if we consider this object to be part of a vortex tangle.
The result (\ref{asym_Gamma}) means that $\Gamma({\sf s}) = 0$ if parameter ${\sf s}$ corresponds to a flex point of a curve ${\boldsymbol r}({\sf s})$. This indicates the stopping  of vortex motion at that location, and, consequently,  effectively disconnection of the initial vortex filament.
Thus, our approach leads us to the number of surprising conclusions. 
\begin{itemize}
{ \item[(a)]  In the proposed model, entangled states  (\ref{Psi_line})  arise with a high degree of probability during turbulence development;}
\item[(b)] Constructed quantum filament-like structures ${\boldsymbol{r}}({\sf{s}})$ naturally admit  circulation values $\Gamma({\sf {s}})$ that are  non-trivial functions of the parameter ${\sf { s}}$. This fact leads  to the emergence of secondary vortices; 
\item[(c)]  In unbounded space, there are no any  straight segments $[{\boldsymbol r}({\sf s}_1), {\boldsymbol r}({\sf s}_2)]$ on a ''vortex filament'' ${\boldsymbol r}({\sf s})$; 
  \item[(d)] Initially, we considered  closed loops ${\boldsymbol r}({\sf s})$  only. 
In the context of the above, we can also consider unclosed filaments. Indeed, assuming that the condition $[R({\sf s})]^{\,-1} \to 0$  is fulfilled at the end of an unclosed filament, we can see that the vortex motion disappears at the filament ends.
Thus, our model allows for the existence of not only closed filaments, but also individual filament fragments. 
\end{itemize}

We believe that the proposed concept of filamentous structures  ${\boldsymbol r}(\sf{s})$ as well as the mechanism of creating and disconnecting these structures inside the vortex tangle,  give a new perspectives  on  the chaos appearance  in  quantum turbulent flow.

\section{Concluding  remarks.}

~~~In this study, we proposed the concept of large vortex-like structures that can arise in quantum turbulent flows. The specific mechanism of disconnection of these objects (vortex filaments) has also been investigated.
In accordance with our theoretical buildings,  the appearance of turbulent flow in a quantum fluid passes through the following stages.
\begin{enumerate}
\item The accidental emergence of a single circular-shaped vortex filament with meso- and micro-scale dimensions;
\item Subsequent decay of this unique vortex, and the emergence of a number of circular-shaped vortex filaments of meso-and micro-scale dimension (see paper \cite{Tal_PoF23}, where the possible scenario of this process was studied);
\item  After the appearance of a sufficient  number of micro vortices due to the decay processes,  the appearance of quantum superpositions (\ref{Psi_line})  and mixed states (\ref{rho_final}) that can be interpreted as  ''large'' vortex filaments ${\boldsymbol r}({\sf s})$; 
\item The emergence of "secondary" vortices near the filament ${\bf r}({s})$ that is due to the difference in the values of the circulation  $\Gamma(s_i)$ for ${\sf s}_1 \neq {\sf s}_2$;
\item The accidental disconnections of large filaments caused by the dynamical appearance of flex-points on a curve ${\boldsymbol r}({\sf s})$. This process results in the emergence of finite and unclosed filament fragments within a quantum vortex tangle;
\item \dots To be continued \dots
\end{enumerate}

 In our opinion, the proposed approach offers a new perspective on understanding such complex phenomena as the appearance of vortex tangles in quantum turbulent flows.

All structures considered were studied in an  unbounded space $\mathbb{R}_3$. 
If we consider the motion of a vortex in a certain bounded domain, then our conclusions must be modified.
This statement is conditioned by the fact that the value of the function $|\beta({\sf s})|$ is bounded because the vortex radius   $R = R({\sf s})$  is bounded. 
Besides, both the spectral problem (\ref{H_quant}) and (\ref{eq_sp_Gamma}) have discrete sets of eigenvalues. Therefore,  the spectra of the circulation $\Gamma$ and the energy $E$ are discrete in the case of a bounded domain.   The specific structure of these spectra provides a new perspective on understanding the origin of quantum turbulence.  We will not discuss this issue in this study. This item was investigated in detail in paper \cite{Tal_Chaos25J}

{ Let us say a few words about the possible experimental verification of the suggested approach.  The experiments for  the circulation measuring  in superfluid helium for a single straight vortex  filament  have been described many years ago \cite{Donn}. It is clear that the  problem  for the quantum turbulent flow is essentially  more complicated.
Any local measurements in a turbulent medium are primarily statistical, as there is no way to exclude the influence of nearby vortices on the measurement. Moreover any local measurements  will be ambiguous here because our system is quantum in nature.
But what is the best way to experimentally prove the theory here?
At this stage, two paths for verifying our theoretical predictions exist in the development of our approach. First, we can verify the conclusions (c) and (d) which are formulated at the end of the previous section. To do this, we need the ability to see an instantaneous picture of the vortex tangle within the flow.  The second direction assumes the further development of our theory and the calculation of a number of observable macroscopic parameters.

Indeed,}
 the mechanism of the connection and  reconnection of considered vortex-like object ${\boldsymbol r}({\sf s})$ must be considered in terms of the Fock space  
\[{\mathfrak H}  =  \bigoplus_{N=0}^{\infty}\boldsymbol{H}_N  \,, \]
where the interaction between the fundamental micro-vortices (see Section 2) can be described qute naturally \cite{Tal_PoF23}.
{ The  development of this approach involves the construction of full Hamiltonian $\widehat{H}$ which describes the dynamics as well as the creation and annihilation of ''primary'' round-shaped micro -and meso- vortices in terms of the space
 ${\mathfrak H}$.  The first steps toward this goal have been taken in the papers \cite{Tal_PoF23,Tal_PhScr}. The next stage of theory development allows us to construct the partition function 
${\cal Z} = {\rm Tr}\,\exp(- \widehat{H}/k_B T)$ and study it. As a consequence, we can calculate thermodynamic functions and the macroscopic characteristics of the described quantum flow, as well as the dependence of these characteristics on the temperature $T$, pressure, external fields   and so on. 
 As it seems, these dependencies can be verified experimentally.
 In this approach, the author considers this method as the only way to compare results with possible experimental data. Undoubtedly, numerical simulations of the processes occurring in  a quantum turbulent flow will be required for the implementation of this program. }
The author hopes to explore this topic in more detail in future studies.

{ \section{ Appendix. The notations used in the  paper.}

\vspace{2mm}

\begin{itemize}
\item[$\widetilde{\mathcal G}_3$~] ~--~ the centrally exteded Galilei group;
\item[$\mu_0$~] ~--~ the central charge for the central extension of the Galilei group;
\item[$R$~] -- the radius of the round-shaped vortex filament;
\item[${\sf a}$~] ~--~ the radius of the filament's core;
\item[$\alpha,\omega$] ~--~ the dimensionless constants that are contained in the equation LIA;
\item[$\varrho_0$] -- the fluid density;
\item[$v_0$] ~--~ the speed of sound in the fluid;
\item[$R_f$] ~--~ the minimal possible value of the radius $R$;
\item[$L$~] ~--~  the characteristic size of the problem (it was explained in the text);
\item[${\cal E}_0 $] ~--~ the energy scale of the problem (it was defined in the text);
\item[$\tilde\mu_0$] ~--~ the auxiliary mass parameter (it was defined in the text);
\item[$t_0$] ~--~ the time scale of the problem (it was defined in the text);
\item[$t$] ~--~ the real (physical) time;
\item[$\xi, \tau $] ~--~ the dimensionless parameters which describe the length and evolution of the filament; 
\item[$t^\#$] ~--~ the conditional time $(t^\# = t_0\tau)$;
\item[$\Phi_\omega $] ~--~ the flow in the filament core;
\item[$\Gamma$~] ~--~ the circulation;
\item[$\boldsymbol{u}$~] ~--~ the fluid velocity outside of a filament;
\item[~$\boldsymbol{w}, \boldsymbol{w}_\parallel,\boldsymbol{w}_\perp $] ~--~ the vorticity vectors for the different objects;
\item[${\boldsymbol{r}}(\sf s)$] ~--~ the smooth curve in the space $\mathbb{R}_3$;
\item[$\sf s$~] ~--~ the natural parameter along the curve ${\boldsymbol{r}}(\sf s)$;
\item[${\boldsymbol{q}}(\sf s)$] ~--~ the evolute of the curve ${\boldsymbol{r}}(\sf s)$;
\item[$\ell$~] ~--~ the natural parameter along the evolute ${\boldsymbol{q}}(\sf s) = {\boldsymbol{q}}(\sf s(\ell))$;
\item[${\mathfrak r}$~] ~--~ the projective radius vector ($ {\boldsymbol{r}}/R$);
\item[$\chi,\varpi$] ~--~ the  variables that describe the radius and core flow of a filament;
\item[$\boldsymbol{H}_1 $] ~--~ the Hilbert space of the quantum system under consideration;
\item[$\widehat{H}^\# $] ~--~ the Hamiltonian that correspond to the conditional time;
\item[$\widehat{H} $~] ~--~ the Hamiltonian that correspond to the real time;
\item[$|\,\beta\rangle $] ~--~ the coherent state for the harmonic oscillator;
\item[$|\, \boldsymbol{z} \rangle $] ~--~ the coherent state for tree-dimensional free particle.

\end{itemize} }

\end{document}